\RequirePackage{ifpdf}
\ifpdf % We are running pdfTeX in pdf mode
\documentclass[pdftex]{sigma}
\else
\documentclass{sigma}
\fi

\begin{document}

\numberwithin{equation}{section}

\newcommand{\bq}{\mathbf{q}}
\newcommand{\bp}{\mathbf{p}}
\newcommand{\bL}{\mathbf{L}}
\newcommand{\ho}{{\rm{O}}}
\newcommand{\KC}{{\rm{KC}}}
\newcommand{\RR}{\mathbb{R}}

\newcommand{\om}{\omega}
\newcommand{\cM}{{\mathcal M}}

\newcommand{\tq}{{\tilde q}}
\newcommand{\tp}{{\tilde p}}

\def\1{\'{\i}}
 \def\dd{{\rm d}}

 \def\la{\lambda}
 \def\te{\theta}
 \def\d{{\rm d}}

  \def\al{\alpha}
  \def\bb{\beta}
 \def\ga{\gamma}
 \def\del{\delta}
 \def\ji{\xi}

 \def\cteb{\beta}

\def\k{\kappa}
\def\>#1{{\mathbf#1}}

\allowdisplaybreaks

\renewcommand{\thefootnote}{$\star$}

\renewcommand{\PaperNumber}{048}

\FirstPageHeading

\ShortArticleName{Superintegrable Oscillator and Kepler Systems on   Spaces of Nonconstant Curvature}

\ArticleName{Superintegrable Oscillator and Kepler Systems \\ on   Spaces of Nonconstant Curvature\\  via the St\"ackel Transform\footnote{This paper is a
contribution to the Special Issue ``Symmetry, Separation, Super-integrability and Special Functions~(S$^4$)''. The
full collection is available at
\href{http://www.emis.de/journals/SIGMA/S4.html}{http://www.emis.de/journals/SIGMA/S4.html}}}

\Author{\'Angel BALLESTEROS~$^\dag$, Alberto ENCISO~$^\ddag$, Francisco J.~HERRANZ~$^\dag$,
Orlando RAGNISCO~$^\S$
  and Danilo RIGLIONI~$^\S$}
\AuthorNameForHeading{\'A.~Ballesteros, A.~Enciso, F.J.~Herranz, O.~Ragnisco and D.~Riglioni}

\Address{$^\dag$~Departamento de F\'{\i}sica, Universidad de Burgos, E-09001
Burgos, Spain}
\EmailD{\href{mailto:angelb@ubu.es}{angelb@ubu.es}, \href{mailto:fjherranz@ubu.es}{fjherranz@ubu.es}}

\Address{$^\ddag$~Instituto de Ciencias Matem\'aticas (CSIC-UAM-UCM-UC3M), Consejo Superior\\
 \hphantom{$^\ddag$}~de Investigaciones Cient\'{\i}ficas, C/ Nicol\'as Cabrera 14-16, E-28049 Madrid,
Spain}
\EmailD{\href{mailto:aenciso@icmat.es}{aenciso@icmat.es}}

\Address{$^\S$~Dipartimento di Fisica,   Universit\`a di Roma Tre and Istituto
Nazionale di Fisica Nucleare\\
 \hphantom{$^\S$}~sezione di Roma Tre,  Via Vasca Navale 84,
I-00146 Roma, Italy}
\EmailD{\href{mailto:ragnisco@fis.uniroma3.it}{ragnisco@fis.uniroma3.it}, \href{mailto:riglioni@fis.uniroma3.it}{riglioni@fis.uniroma3.it}}

\ArticleDates{Received March 18, 2011, in f\/inal form May 12, 2011;  Published online May 14, 2011}

\Abstract{The St\"ackel transform is applied to the geodesic motion on Euclidean space, through the harmonic oscillator and Kepler--Coloumb potentials, in order to obtain ma\-ximal\-ly superintegrable classical systems on $N$-dimensional Riemannian spaces of nonconstant curvature. By one hand, the harmonic oscillator potential leads to two families of superintegrable systems which are interpreted as an intrinsic Kepler--Coloumb  system on a hyperbolic curved space and as the so-called  Darboux~III oscillator. On the other, the    Kepler--Coloumb potential gives rise to   an oscillator system on a spherical curved space as well as   to the Taub-NUT oscillator. Their  integrals of motion are explicitly given.
The role of the   (f\/lat/curved)  Fradkin tensor and   Laplace--Runge--Lenz $N$-vector for all of these Hamiltonians  is highlighted throughout   the paper. The corresponding quantum maximally superintegrable systems are also presented.}

\Keywords{coupling constant metamorphosis;  integrable systems;  curvature; harmonic
oscillator; Kepler--Coulomb; Fradkin tensor;  Laplace--Runge--Lenz vector; Taub-NUT; Darboux surfaces}

\Classification{37J35; 70H06; 81R12}

\section{Introduction}

The coupling constant metamorphosis or St\"ackel transform was formerly introduced in~\cite{Hietarinta, SIAM} and  further developed and applied to several classical and quantum Hamiltonian systems  in~\cite{Stackel2,Stackel4,Sergyeyev, Daska, Stackel5}. This approach has proven to be a useful tool in order to relate dif\/ferent (super)integrable systems together with their associated symmetries and to deduce new integrable Hamiltonian systems starting from known ones.

For our purposes, the classical St\"ackel transform can be brief\/ly summarized  as follows~\cite{Stackel2,Stackel4}.
Consider the  conjugate coordinates and momenta $\bq,\bp\in\RR^N$   with canonical Poisson  bracket $\{q_i,p_j\}=\delta_{ij}$ and the notation:
\[
\bq^2=\sum_{i=1}^N q_i^2,\qquad \bp^2=\sum_{i=1}^N p_i^2,\qquad |\bq|=\sqrt{\bq^2}  .
\]
Let $H$ be an ``initial'' Hamiltonian, $H_U$ an ``intermediate'' one  and ${\tilde H}$ the ``f\/inal'' system given by
\begin{gather}
H=\frac{\bp^2}{\mu(\bq)}+V(\bq),\qquad
H_U=\frac{\bp^2}{\mu(\bq)}+U(\bq),\qquad
 \tilde H=\frac{H}{U}=\frac{\bp^2}{\tilde\mu(\bq)}+ \tilde V(\bq),
\label{ya}
\end{gather}
such that   $\tilde \mu=\mu U $ and $ \tilde V= {V}/{U}$. Then, each {\em second-order} integral of motion   (symmetry)~$S$   of~$H$ leads to a new one~$\tilde S$ corresponding to $\tilde H$ through an ``intermediate''  symmetry $S_U$ of $H_U$. In particular, if~$S$ and~$S_U$ are written as
\begin{gather}
S=\sum_{i,j=1}^Na^{ij}(\bq)p_ip_j+W(\bq)=S_0+W(\bq)
,\qquad S_U=S_0+W_U(\bq),
\label{ye}
\end{gather}
then one gets  a second-order symmetry of $\tilde H$ in the form
 \begin{gather}
 \tilde S=S_0- {W_U} \tilde H  .
 \label{yf}
 \end{gather}

The aim of this paper is to apply the above procedure when the ``initial'' Hamiltonian $H$ is the  $N$-dimensional ($N$D)  free Euclidean motion, and when $H_U$ is either the isotropic harmonic oscillator or the Kepler--Coulomb (KC) Hamiltonian. It is well known that these three systems are maximally superintegrable (MS), that is, they are endowed with the maximum number of $2N-1$ functionally  independent integrals of motion (in these cases, all of them are quadratic in the momenta). These three systems and their MS  property  are brief\/ly recalled in the next section, and we will see that the St\"ackel transform gives rise to
several  MS systems ${\tilde H}$ that are def\/ined on Riemannian spaces of {\em nonconstant curvature}. Moreover, we will show that  the new potential $\tilde V$ can be interpreted as either an (intrinsic) oscillator or a KC potential on the corresponding  curved manifold. In this way, by starting from the Euclidean Fradkin tensor~\cite{Fradkin}, formerly studied  by Demkov  in~\cite{Demkov}, and the Laplace--Runge--Lenz (LRL) $N$-vector,    the St\"ackel transform provides for each case its curved analogue (see~\cite{BEHR09, Kepler, commun2} and references therein).

In particular, we show in Section~\ref{section3} that if $H_U$ is chosen to be the harmonic oscillator we obtain two dif\/ferent f\/inal MS Hamiltonians, for which  $\tilde H$ is  endowed with a curved Fradkin tensor; these are  a KC system on a   hyperbolic space
of nonconstant curvature and the so-called Darboux III oscillator~\cite{PhysD, annals, Gadella}. In Section~\ref{section4} we take $H_U$ as the (f\/lat) KC Hamiltonian and the St\"ackel transform leads to other two  dif\/ferent  MS systems together with their curved LRL $N$-vector; both of them are interpreted as intrinsic oscillators on   curved Riemannian manifolds. Surprisingly enough, one of them is the $N$D generalization of the Taub-NUT oscillator~\cite{GM86, FH87, GR88, IK94, IK95, BCJ, JL, GW07}.
We stress that for some systems the dimension $N=2$ is rather special as the underlying manifold remains f\/lat, meanwhile for $N\ge 3$ such systems are def\/ined on proper curved spaces (see Sections~\ref{section3.1} and~\ref{section4.1}). This    is similar to what happens in the  classif\/ications of 2D and 3D  integrable systems on spaces of constant curvature (including the f\/lat Euclidean one)~\cite{Evans, 12, 20, 19, Santander, 21, 18, car2} which exhibit  some dif\/ferences according to the dimension and, in general,  the 3D case is  usually the cornerstone for the generalization of a given system to arbitrary dimension.

As a byproduct of this construction, the ``growth'' of the
Fradkin tensor and the LRL vector from their Euclidean ``seeds''   to their curved counterparts can be highlighted from a global perspective.  These results are comprised in Table~\ref{table1} in  the last section. Furthermore we also present   in Table~\ref{table2}  the MS quantization for all of the above systems together with their ``additional'' quantum Fradkin/LRL symmetries.

\section{Harmonic oscillator and  Kepler potentials on Euclidean space}\label{section2}

In order to f\/ix a suitable common framework, we  brief\/ly recall the well-known  basics of the superintegrability properties of the Hamiltonians describing free motion, harmonic  oscillator and KC potentials on the $N$D Euclidean space ${\bf E}^N$.

As the ``initial'' Hamiltonian $H$ (\ref{ya}) for the St\"ackel procedure we consider the one def\/ining the geodesic motion on  ${\bf E}^N$ plus a {\em relevant} constant $\al$:
\begin{gather}
H=\frac 12\, \bp^2 +\al .
\label{ba}
\end{gather}
Obviously this is a MS system, and there are many possibilities to choose its integrals of motion. We
shall make use, throughout   the   paper,  of the  following results.

\begin{proposition}\label{proposition1}\quad
\begin{enumerate}\itemsep=0pt
\item[$(i)$] The  Hamiltonian \eqref{ba} is endowed with the following constants of motion.
\begin{itemize}\itemsep=0pt
\item $(2N-3)$  angular momentum integrals $(m=2,\dots,N)$:
\begin{gather}
  S^{(m)}= \sum_{1\leq i<j\leq m}   (q_ip_j-q_jp_i)^2 , \qquad
 S_{(m)}=  \sum_{N-m<i<j\leq N}   (q_ip_j-q_jp_i)^2 , \nonumber\\  S^{(N)}=S_{(N)}\equiv  \bL^2,  \label{bb}
 \end{gather}
  where $ \bL^2$ is the square of the  total angular momentum.

\item $N^2$ integrals  which are the ``seeds'' of the Fradkin tensor  $(i,j=1,\dots,N)$:
 \begin{gather}
 S_{ij}=p_ip_j  \qquad \mbox{such that}\quad        \sum_{i=1}^N S_{ii} = 2(H-\al) .
\label{bc}
\end{gather}

\item $N$ integrals  which are the  ``seeds'' of the components of  the LRL  vector   $(i=1,\dots,N)$:
 \begin{gather}
 S_{i}=\sum_{k=1}^N p_k \left( q_k p_i - q_i p_k \right)  \qquad \mbox{such that}\quad
 \sum_{i=1}^N S_{i}^2= 2 \bL^2 (H-\al).\label{bd}
\end{gather}
\end{itemize}

\item[$(ii)$] Each of the three  sets $\{ H,S^{(m)}\}$,
$\{H,S_{(m)}\}$ $(m=2,\dots,N)$ and   $\{S_{ii}\}$ $(i=1,\dots,N)$ is  formed by $N$ functionally independent functions  in involution.

\item[$(iii)$] Both sets $\{ H,S^{(m)}, S_{(m)},  S_{ii} \}$  and $\{ H,S^{(m)}, S_{(m)},  S_{i} \}$     $(m=2,\dots,N$ and a fixed index $i)$    are  constituted  by $2N-1$ functionally independent functions.
    \end{enumerate}
\end{proposition}

As ``intermediate'' Hamiltonians $H_U$ (\ref{ya}) we consider either the harmonic oscillator or the KC system.
Since both of them are central potentials,  the angular momentum integrals (\ref{bb}) are valid
  for both cases, that is, $  S^{(m)}_U\equiv  S^{(m)}$ and   $S_{U,{(m)}}\equiv  S_{(m)}$ in (\ref{ye}). We recall that, in fact, the spherical symmetry of a central potential on  ${\bf E}^N$ directly provides such  $(2N-3)$    independent angular momentum integrals, so they characterize a
    quasi-MS system~\cite{BH07, sigmaorlando}. However what makes rather special the  harmonic oscillator and KC systems is the existence of one more independent integral, which is extracted from a new set of integrals that ensure their MS property and is related to the fact that these two systems are the only ones fulf\/illing the classical Bertand's theorem~\cite{Be73}. In this respect, each of the sets of integrals (\ref{bc}) and (\ref{bd}) gives rise to one known set of additional constants for the harmonic oscillator and KC system, respectively.

\begin{proposition}\label{proposition2} \qquad
\begin{enumerate}\itemsep=0pt

\item[$(i)$] The  harmonic oscillator Hamiltonian defined by
\begin{gather}
H_U=\frac 12  \bp^2+\bb \bq^2 +\ga
\label{be}
\end{gather}
has the  $(2N-3)$  angular momentum integrals \eqref{bb} together
 with  $N^2$ additional ones  given by the components of  the   ND  Fradkin tensor  $(i,j=1,\dots,N)$:
 \begin{gather*}
 S_{U,{ij}}=p_ip_j  + 2\bb q_i q_j\qquad \mbox{such that}\quad        \sum_{i=1}^N S_{U,{ii} }= 2(H_U-\ga) .
%\label{bf}
\end{gather*}

\item[$(ii)$] Each of the three  sets $\{ H_U,S^{(m)}\}$,
$\{H_U,S_{(m)}\}$ $(m=2,\dots,N)$ and   $\{S_{U,{ii}}\}$ $(i=1,\dots,N)$ is  formed by $N$ functionally independent functions  in involution.

\item[$(iii)$] The set $\{ H_U,S^{(m)}, S_{(m)},  S_{U,{ii}} \}$       $(m=2,\dots,N$ and a fixed index $i)$    provides  $2N-1$ functionally independent functions.
\end{enumerate}
\end{proposition}

\begin{proposition}\label{proposition3}\qquad
\begin{enumerate}
\item[$(i)$] The KC Hamiltonian given by
\begin{gather}
H_U=\frac 12  \bp^2+\frac{\del}{| \bq|} +\ji
\label{bg}
\end{gather}
has the  $(2N-3)$  angular momentum integrals \eqref{bb} together
 with  the $N$   components of  the      LRL vector  $(i=1,\dots,N)$:
 \begin{gather*}
 S_{U,{i}}=\sum_{k=1}^N p_k \left( q_k p_i - q_i p_k \right) -\frac{\del q_i}{|\bq|} \qquad \mbox{such that}\quad      \sum_{i=1}^N S_{U,{i}}^2= 2 \bL^2 (H_U-\ji)+\del^2 .
%\label{bh}
\end{gather*}

\item[$(ii)$] Each of the two  sets $\{ H_U,S^{(m)}\}$ and
$\{H_U,S_{(m)}\}$ $(m=2,\dots,N)$   is  formed by $N$ func\-tional\-ly independent functions  in involution.

\item[$(iii)$] The set $\{ H_U,S^{(m)}, S_{(m)},  S_{U,{i}} \}$       $(m=2,\dots,N$ and a fixed index $i)$    is  constituted  by $2N-1$ functionally independent functions.
\end{enumerate}
\end{proposition}

In the two next sections we apply the St\"ackel transform to each of   these two MS systems. Notice that the proper isotropic harmonic oscillator arises whenever $\bb=\omega^2/2$ with frequency~$\om$ and $\ga=0$, while the Kepler one corresponds to set $\del=-K$ and $\ji=0$.
 We remark that in this approach the constant $\al$
is essential in order to obtain a curved potential while the others  $\bb$, $\ga$, $\del$ and $\ji$ enter in both the kinetic and the potential term  giving rise to MS  oscillator/KC potentials on Riemannian spaces of nonconstant curvature, so that they can be regarded as    classical ``deformation parameters''.

\section{Superintegrable systems from harmonic oscillator potential}\label{section3}

If   we consider  as  the  initial Hamiltonian $H$ the free system  (\ref{ba}) and as  the intermediate one the harmonic oscillator $H_U$ (\ref{be}), then we obtain the f\/inal Hamiltonian $ {\tilde H}$
\begin{gather}
 {\tilde H} =   \frac{\bp^2}{2(\ga+\bb \bq^2)}   +\frac{\al}{\ga+\bb \bq^2} ,
 \label{ca}
\end{gather}
so that the relations  (\ref{ya}) read as
\[
 \mu=2,\qquad V=\alpha,\qquad
 U=\ga+\bb\bq^2,\qquad  \tilde \mu=2\big(\ga+\bb \bq^2\big),\qquad \tilde V=\frac{\alpha}{\ga+\bb \bq^2} .
% \label{cb}
\]
As far as the symmetries $S=S_0+W$   (\ref{ye}) are concerned, we f\/ind from Proposition~\ref{proposition1} that
   \begin{gather*}
  S_0^{(m)}=S^{(m)},\qquad W^{(m)}=0,\qquad   S_{0,{(m)}}=S_{(m)},\nonumber\\ W_{(m)}=0,\qquad   S_{0,ij}=S_{ij},\qquad W_{ij}=0,
  %\label{cc}
 \end{gather*}
while from Proposition~\ref{proposition2} we obtain the elements $W_U$ for the decompositions of $S_U=S_0+W_U$,
\begin{gather*}
  W_U^{(m)}=0,\qquad  W_{U,(m)}=0,\qquad W_{U,ij}=2\bb q_i q_j ,
 % \label{cd}
 \end{gather*}
where $m=2,\dots,N$ and $i,j=1,\dots, N$.

Consequently, the Hamiltonian $ {\tilde H}$ (\ref{ca})  is St\"ackel equivalent to the free Euclidean motion, through the harmonic oscillator potential, and
  its integrals of motion $\tilde S$ come from  (\ref{yf}) and turn out to be
 \begin{gather}
  \tilde S^{(m)}   =S^{(m)}   ,\qquad
  \tilde S_{(m)}   =S_{(m)}    , \qquad
  \tilde S_{ij}=p_ip_j -2\bb q_iq_j\tilde H (\bq,\bp)  .
  \label{ce}
 \end{gather}
 Thus we have obtained the
 $(2N-3)$ angular momentum integrals $  S^{(m)} $ and $  S_{(m)} $, together with~$N^2$ ones, $ \tilde S_{ij}$, which form a {\em curved} Fradkin tensor.   The quasi-MS property of $\tilde H$ is ensured by the preservation of the  $(2N-3)$ angular momentum integrals,
 that is, each of the two  sets $\{ \tilde H,S^{(m)}\}$,
$\{\tilde H,S_{(m)}\}$ ($m=2,\dots,N$)  is  formed by $N$ functionally independent functions  in involution.
Hence, from now on, we assume this fact and only pay attention to the additional  constants $ \tilde S_{ij}$
 which characterize (\ref{ca})  as a MS system.

  In order to perform a preliminary  geometrical analysis of   $ {\tilde H} $ we recall that, in general,
 any Hamiltonian of the form
\[
 {\cal H} =\frac{\bp^2}{2f(|\bq|)^2}+{\cal V}(|\bq|)
 \label{cf}
\]
 can be interpreted as   describing a particle (with unit mass) on an $N$D spherically symmetric space $\cM$
 under the action of the central potential ${\cal V}(|\bq|)$~\cite{annals}. The   metric  and scalar curvature of $\cM$ are given by~\cite{PLB}
 \begin{gather}
 \dd s^2=f(|\bq|)^2 \dd\bq^2,\nonumber\\
  R=-(N-1)\left( \frac{    (N-4)f'(r)^2+  f(r)  \left(    2f''(r)+2(N-1)r^{-1}f'(r)  \right)}   {f(r)^4  } \right) ,
 \label{cg}
\end{gather}
 where we have introduced the radial coordinate $r=|\bq|$. For general results on 2D and 3D  (super)integrable systems on conformally f\/lat spaces   we refer to~\cite{darboux3, darboux5, darboux7}.

 Furthermore, the conformal factor $f(|\bq|)=f(r)$ is directly related, under
 the following prescription, with  the   {\em intrinsic KC and oscillator  potentials}  on   $\cM$:
 \begin{gather}
{\cal U}_\KC(r) : =\int^r\frac{\dd r'}{r'^2f(r')} ,\qquad {\cal U}_\ho(r) :=\frac  1{ {\cal U}_\KC(r)^2} ,
\label{ch}
  \end{gather}
that was introduced in~\cite{annals}  up to additive and multiplicative constants.

  With these ideas in mind, we now analyze the specif\/ic systems def\/ined by $\tilde H$ (\ref{ca}) according to the values of the parameters
  $\bb$ and $\ga$.  Notice that $\al$ is the constant which governs the potential, so to setting $\al=0$ leads to geodesic motion on $\cM$, and that $\bb$ must be always dif\/ferent from zero, since otherwise $\tilde H$ is again the initial $H$. Therefore we are led to consider two dif\/ferent cases with generic $\al$: $(i)$ $\bb\ne 0$, $\ga=0$; and $(ii)$  $\bb\ne 0$, $\ga\ne0$.

\subsection[The case with $\bb\ne 0$ and $\ga=0$: a curved hyperbolic KC system]{The case with $\boldsymbol{\bb\ne 0}$ and $\boldsymbol{\ga=0}$: a curved hyperbolic KC system}\label{section3.1}

If $\ga=0$ we scale $\tilde H$ to deal with the Hamiltonian
\begin{gather}
{\cal H}_\KC=\beta \tilde H=   \frac{\bp^2}{  2\bq^2}  +\frac{\al}{ \bq^2} .
\label{da}
\end{gather}
Then, $f(|\bq|)=|\bq| =r$ so  the   metric and scalar curvature (\ref{cg}) on $\cM$ reduces to
\begin{gather}
\dd s^2=\bq^2 \dd\bq^2,\qquad
  R=-\frac{3(N-1)(N-2)}{r^4},
  \label{db}
\end{gather}
while the  intrinsic KC and oscillator  potentials (\ref{ch}) on this curved space would be
\begin{gather*}
{\cal U}_\KC(r)=-\frac{1}{2 r^2} ,\qquad {\cal U}_\ho(r)= 4 r^4.
%\label{dc}
\end{gather*}
The latter result shows that ${\cal H}_\KC$ (\ref{da}) always def\/ines an intrinsic KC potential on the space~$\cM$. Nevertheless the curvature (\ref{db}) vanishes for $N=2$, while the space is of nonconstant curvature for $N\ge 3$. Therefore, for $N=2$ the Hamiltonian must correspond to the usual KC system on the Euclidean space. This fact can be explicitly proven by applying   to   ${\cal H}_\KC$ (\ref{da}) the Kustaanheimo--Stiefel canonical transformation def\/ined by~\cite{Kustaanheimo1,Kustaanheimo2}
\begin{gather}
 \tq_1=\frac12 \big(q_1^2 - q_2^2\big),\qquad \tp_1= \frac{p_1 q_1 - p_2 q_2}{q_1^2 + q_2^2} , \qquad
 \tq_2=q_1 q_2,\qquad \tp_2= \frac{p_2 q_1 +p_1 q_2}{q_1^2 + q_2^2} ,
\label{za}
\end{gather}
so with canonical Poisson bracket $\{\tq_i,\tp_j\}=\delta_{ij}$. In this way we recover  the  2D KC Hamiltonian
\begin{gather*}
{\cal H}_\KC=      \frac{p_1^2+p_2^2}{2(q_1^2+q_2^2)}+\frac{\al}{ {q_1^2+q_2^2}}=     \frac 12\big(\tp_1^2+\tp_2^2\big)+\frac{\al}{ 2\sqrt{\tq_1^2+\tq_2^2}}
%\label{zb}
\end{gather*}
and the {\em five}  symmetries (\ref{ce}) reduce to {\em three} integrals of motion, namely
\begin{gather*}
   S^{(2)}=  S_{(2)}=\bL^2=(q_1 p_2 - q_2 p_1)^2= 4 (\tq_1 \tp_2 - \tq_2 \tp_1)^2,\nonumber\\
 \tilde S_{11}=p_1^2 - 2 q_1^2{\cal H}_\KC=2 \tp_2(\tq_2\tp_1- \tq_1\tp_2)-\frac{\al\tq_1}{\sqrt{\tq_1^2+\tq_2^2}}-\al ,
\qquad \tilde S_{22}=-\tilde S_{11}-2\al,
 \nonumber \\
  \tilde S_{12}=\tilde S_{21}=p_1p_2- 2 q_1 q_2{\cal H}_\KC=2 \tp_1(\tq_1\tp_2- \tq_2\tp_1)-\frac{\al\tq_2}{\sqrt{\tq_1^2+\tq_2^2}} .%\label{zc}
\end{gather*}
Hence, by taking into account Proposition~\ref{proposition3} for $N=2$, we f\/ind that, under the above canonical transformation, the only angular momentum integral $ S^{(2)}$ is   kept, while the four constants coming from the 2D Fradkin tensor reduce to the two components of the LRL  vector: $(\tilde S_{11},\tilde S_{22})\to \tilde S_1$ and  $(\tilde S_{12},\tilde S_{21})\to \tilde S_2$.

Consequently, a proper curved KC system arises whenever $N\ge 3$ and its full integrability properties can be summarized as follows.

\begin{proposition}\label{proposition4}\qquad
\begin{enumerate}\itemsep=0pt
\item[$(i)$] For $N\ge 3$, the   Hamiltonian ${\cal H}_\KC$ \eqref{da} determines an intrinsic KC system on the hyperbolic space of nonconstant curvature~\eqref{db}.

\item[$(ii)$] ${\cal H}_\KC$ is endowed with the  $(2N-3)$  angular momentum integrals \eqref{bb} together
 with  $N^2$ ones  which are the components of  an   ND  curved Fradkin tensor  $(i,j=1,\dots,N)$:
 \begin{gather*}
 \tilde S_{ij}=p_ip_j  -2  q_i q_j {\cal H}_\KC \qquad \mbox{such that}\quad        \sum_{i=1}^N \tilde S_{ii} = - 2\al\quad \mbox{and}\quad \{ \tilde S_{ii}, \tilde S_{jj} \}=0 .
%\label{bff}
\end{gather*}

\item[$(iii)$] The set $\{ {\cal H}_\KC,S^{(m)}, S_{(m)},  \tilde S_{{ii}} \}$       $(m=2,\dots,N$ and a fixed index $i)$    is formed by  $2N-1$ functionally independent functions.
\end{enumerate}
\end{proposition}

\subsection[The case with $\bb\ne 0$ and $\ga\ne0$: the Darboux III oscillator]{The case with $\boldsymbol{\bb\ne 0}$ and $\boldsymbol{\ga\ne0}$: the Darboux III oscillator}\label{section3.2}

If both  $\bb,\ga\ne0$, we can write the Hamiltonian (\ref{ca})  in the form
\begin{gather}
{\cal H}_\la =\ga {\tilde H}-\al=\frac{\bp^2}{2(1+\la \bq^2)}-\frac{\la \al\, \bq^2}{1+\la \bq^2},\qquad \la=\bb/\ga .
\label{df}
\end{gather}
Then the   metric and scalar curvature (\ref{cg}) on the corresponding manifold  $\cM$   are given by
\begin{gather}
\dd s^2=\big(1+\la \bq^2\big)\dd\bq^2,\qquad
  R=-\la\,\frac{(N-1)\bigl( 2N+3(N-2)\la r^2\bigr)}{(1+\la r^2)^3},
  \label{dg}
\end{gather}
and the  intrinsic   potentials (\ref{ch}) read
\begin{gather}
{\cal U}_\KC(r)=-\frac{\sqrt{1+\la r^2}}{r} ,\qquad {\cal U}_\ho(r)= \frac{r^2} { {1+\la r^2}} .
\label{dh}
\end{gather}
In this way, we recover the $N$D spherically symmetric generalization of the Darboux surface of type III~\cite{Ko72,KKMW03, pogo1, pogo2}  introduced in~\cite{annals,PLB}. Notice that the domain of $r=|\bq|$ and the type of the underlying curved manifold depends on the sign of $\la$~\cite{Gadella}:
\begin{gather*}
\la>0:\quad  R(0)=-2\la N(N-1),\qquad r\in [0,\infty); \\
\la<0:\quad R(0)=2|\la| N(N-1),\qquad\, r\in [0,1/\sqrt{|\la|}),
\end{gather*}
where we have written the value of the scalar curvature (\ref{dg}) at the origin $r=0$. We stress that $R(0)$  coincides either with  the scalar curvature of the $N$D {\em hyperbolic space} with negative constant sectional curvature equal to $-2\la$ for $\la>0$, or  with   that corresponding to the $N$D {\em spherical space} with sectional curvature equal to $2|\la|$ for $\la<0$.

 By taking into account the above geometrical considerations and
  expressions~(\ref{dh}), we f\/ind  that ${\cal H}_\la$ comprises  both an intrinsic   hyperbolic oscillator potential and a spherical one on $\cM$ according to the sign of $\la$. Strictly speaking the curved oscillator potentials arise by introducing the frequency $\om^2=-2\la\al $ and, in that form, the limit $\la \to 0$ gives rise to the harmonic oscillator on ${\bf E}^N$, so $\la$ behaves as a classical deformation parameter governing the curvature and the potential. The MS property of~ ${\cal H}_\la$ is then characterized by~\cite{PhysD, Gadella}:

\begin{proposition}\label{proposition5}\qquad
\begin{enumerate}\itemsep=0pt
\item[$(i)$] The   Hamiltonian ${\cal H}_\la$ \eqref{df} defines an intrinsic curved hyperbolic Darboux oscillator for  $\la>0$ and $r\in [0,\infty)$ and a curved spherical Darboux  one for $\la<0$ and $r\in [0,1/\sqrt{|\la|})$.

\item[$(ii)$] Besides the  $(2N-3)$  angular momentum integrals \eqref{bb},   ${\cal H}_\la$ Poisson-commutes with
 the~$N^2$  components of  the   ND  curved Fradkin tensor  $(i,j=1,\dots,N)$ given by
 \begin{gather*}
 \tilde S_{ij}=p_ip_j  -2\la  q_i q_j \left( {\cal H}_\la+\al \right) \qquad \mbox{such that}\quad        \sum_{i=1}^N \tilde S_{ii} =  2 {\cal H}_\la
 \quad \mbox{and}\quad \{ \tilde S_{ii}, \tilde S_{jj} \}=0 .
%\label{di}
\end{gather*}

\item[$(iii)$] The set $\{ {\cal H}_\la,S^{(m)}, S_{(m)},  \tilde S_{{ii}} \}$       $(m=2,\dots,N$ and a fixed index $i)$    is formed by  $2N-1$ functionally independent functions.
\end{enumerate}
\end{proposition}

\section[Superintegrable systems from the Kepler-Coulomb potential]{Superintegrable systems from the Kepler--Coulomb potential}\label{section4}

In this case we consider   the   initial Hamiltonian $H$   (\ref{ba}) and the KC system $H_U$ (\ref{bg}) for
   the intermediate one; this provides the f\/inal Hamiltonian $ {\tilde H}$ (\ref{ya})
\begin{gather}
 {\tilde H} =  \frac{|\bq| \bp^2}{2(\del+\ji |\bq|)}   +\frac{\alpha|\bq|}{\del+\ji |\bq|}   ,
 \label{ea}
\end{gather}
where
\[
 \mu=2,\qquad V=\alpha,\qquad
 U=\frac{\del+\ji |\bq|}{|\bq|}  ,\qquad  \tilde \mu=\frac{2(\del+\ji |\bq|)}{|\bq|} ,\qquad \tilde V=\frac{\alpha|\bq|}{\del+\ji |\bq|} .
 %\label{eb}
\]
From Proposition~\ref{proposition1}  we obtain the decomposition of the symmetries $S=S_0+W$ (\ref{bb}) and (\ref{bd}) ($m=2,\dots,N$ and $i=1,\dots, N$):
\begin{gather*}
  S_0^{(m)}=S^{(m)},\qquad W^{(m)}=0,\qquad   S_{0,{(m)}}=S_{(m)},\qquad W_{(m)}=0,\qquad   S_{0,i}=S_{i},\qquad W_{i}=0,
  %\label{ec}
\end{gather*}
while   from Proposition~\ref{proposition3} we f\/ind the one corresponding  to $S_U=S_0+W_U$  (\ref{ye})
\begin{gather*}
  W_U^{(m)}=0,\qquad  W_{U,(m)}=0,\qquad W_{U,i}=-\frac{\del q_i}{|\bq|}.
  %\label{ed}
\end{gather*}
 Therefore,   the Hamiltonian $ {\tilde H}$ (\ref{ea})  is St\"ackel equivalent to the free Euclidean motion, through the KC potential,  and   its integrals of motion $\tilde S$   (\ref{yf}) are given by
 \begin{gather}
  \tilde S^{(m)}   =S^{(m)}   ,\qquad
  \tilde S_{(m)}   =S_{(m)}    , \qquad
  \tilde S_{i}=\sum_{k=1}^N p_k \left( q_k p_i - q_i p_k \right)+\frac{\del q_i}{|\bq|} \tilde H (\bq,\bp)  .
  \label{ee}
\end{gather}

 Hence,  $ {\tilde H}$ (\ref{ea})  is endowed with the
 $(2N-3)$ angular momentum integrals (\ref{bb}) together with a  {\em curved}  LRL $N$-vector with components $  \tilde S_{i}$.

  Notice that the parameter $\del$  cannot vanish in order to avoid the initial system $H$. Thus, similarly to the previous section, we   study   two systems covered by $ {\tilde H}$ (\ref{ea}):   $(i)$ $\del\ne 0$, $\ji=0$; and $(ii)$  $\del\ne 0$, $\ji\ne0$.

\subsection[The case with $\del\ne 0$ and $\ji=0$:  a curved spherical oscillator system]{The case with $\boldsymbol{\del\ne 0}$ and $\boldsymbol{\ji=0}$:  a curved spherical oscillator system}\label{section4.1}

If $\ji =0$ we  have the Hamiltonian system
\begin{gather}
{\cal H}_\ho=\del \tilde H=\frac 12  |\bq|  \bp^2  + {\al}{| \bq|} .
\label{fa}
\end{gather}
We stress that for $N=3$ this system was early considered in~\cite{SIAM}.
 The   metric and scalar curvature~(\ref{cg})   give
\begin{gather}
\dd s^2=\frac{1}{|\bq|} \dd\bq^2,\qquad
  R=\frac{3(N-1)(N-2)}{4r},
  \label{fb}
\end{gather}
and the  intrinsic KC and oscillator  potentials~(\ref{ch})  turn out to be
\begin{gather*}
{\cal U}_\KC(r)=-\frac{2}{\sqrt{ r}} ,\qquad {\cal U}_\ho(r)= \frac{r}4.
%\label{fc}
\end{gather*}
Hence  ${\cal H}_\ho$ (\ref{fa})  determines   an intrinsic oscillator potential on   $\cM$.
However, for $N=2$ the curvature is equal to zero, so this case should actually be the 2D harmonic oscillator. This can be proven by means of the canonical transformation
\begin{alignat*}{3}
&\tq_1=\frac{q_2}{\left( \sqrt{q_1^2+q_2^2}-q_1\right)^{1/2}},\qquad &&
\tp_1=\frac{ \left( p_1 q_2 - 2 p_2 q_1\right) \left( \sqrt{q_1^2+q_2^2}-q_1\right) +p_2 q_2^2}{\left( \sqrt{q_1^2+q_2^2}-q_1\right)^{3/2}},& \nonumber \\
&\tq_2=\left(\sqrt {q_1^2+q_2^2}-q_1\right)^{1/2},\qquad &&  \tp_2=\frac{   p_2 q_2 - p_1  \left( \sqrt{q_1^2+q_2^2}-q_1\right)  }{\left( \sqrt{q_1^2+q_2^2}-q_1\right)^{1/2}}, &
%\label{fd}
\end{alignat*}
which is just the inverse of the   canonical transformation (\ref{za});  this yields  the expected system
\begin{gather*}
{\cal H}_\ho=  \frac 12 \sqrt{q_1^2+q_2^2}   \big( p_1^2+p_2^2\big) + {\al}{\sqrt {q_1^2+q_2^2}}=     \frac 14\big(\tp_1^2+\tp_2^2\big)+\frac 12  {\al}   \big(\tq_1^2+\tq_2^2\big)  .
%\label{fe}
\end{gather*}
 The canonical transformation of the {\em three} symmetries (\ref{ee}) gives
\begin{gather*}
  S^{(2)}=  S_{(2)}=\bL^2=(q_1 p_2 - q_2 p_1)^2= \frac 14 (\tq_1 \tp_2 - \tq_2 \tp_1)^2,\nonumber\\
\tilde S_{1}= p_2(q_2p_1- q_1p_2)+\frac{ q_1}{\sqrt{q_1^2+q_2^2}}\,
{\cal H}_\ho = \frac 14 \big(\tp_1^2 -\tp_2^2\big)+\frac 12\al\big(\tq_1^2-\tq_2^2\big),\nonumber\\
 \tilde S_{2}=p_1(q_1p_2- q_2p_1)+\frac{ q_2}{\sqrt{q_1^2+q_2^2}}\,
{\cal H}_\ho =\frac 12 \tp_1\tp_2 +\al \tq_1 \tq_2 .%\label{ff}
\end{gather*}
Then the four components of the 2D  Euclidean Fradkin tensor $\tilde S_{ij}$  are recovered, in the new canonical variables, from  the set of constants $({\cal H}_\ho,  \tilde S_{1}, \tilde S_{2})$
by setting
\begin{gather*}
\tilde S_{11}=2 ({\cal H}_\ho+ \tilde S_{1})=\tp_1^2+ 2\al \tq_1^2,
\qquad  \tilde S_{22}=2 ({\cal H}_\ho- \tilde S_{1})=\tp_2^2+ 2\al \tq_2^2,\nonumber\\
 \tilde S_{12}=\tilde S_{21}=
2 \tilde S_{2}=\tp_1\tp_2 +2 \al \tq_1 \tq_2 .
\end{gather*}

Therefore the proper curved system arises whenever $N\ge 3$, which yields the following

\begin{proposition}\label{proposition6}\qquad
\begin{enumerate}\itemsep=0pt
\item[$(i)$] For $N\ge 3$, the   Hamiltonian ${\cal H}_\ho$ \eqref{fa} defines an intrinsic oscillator potential on the spherical space of nonconstant curvature \eqref{fb}.

\item[$(ii)$] ${\cal H}_\ho$ Poisson-commutes with the   $(2N-3)$  angular momentum integrals \eqref{bb} and with
 the components of the LRL $N$-vector given by   $(i=1,\dots,N)$:
 \begin{gather*}
  \tilde S_{i}=\sum_{k=1}^N p_k \left( q_k p_i - q_i p_k \right)+\frac{  q_i}{|\bq|} {\cal H}_\ho  \qquad \mbox{such that}\quad        \sum_{i=1}^N \tilde S_{i}^2 ={\cal H}_\ho^2 - 2\al \bL^2 .
%\label{fi}
\end{gather*}

\item[$(iii)$] The set $\{ {\cal H}_\ho,S^{(m)}, S_{(m)},  \tilde S_{{i}} \}$       $(m=2,\dots,N$ and a fixed index $i)$    is formed by  $2N-1$ functionally independent functions.
\end{enumerate}
\end{proposition}

\subsection[The case with $\del\ne 0$ and $\ji\ne 0$: the Taub-NUT oscillator]{The case with $\boldsymbol{\del\ne 0}$ and $\boldsymbol{\ji\ne 0}$: the Taub-NUT oscillator}\label{section4.2}

We scale the Hamiltonian (\ref{ea})  as
\begin{gather}
 {\cal H}_\eta = \ji {\tilde H}= \frac{|\bq| \bp^2}{2(\eta+  |\bq|)}   +\frac{\alpha|\bq|}{\eta+  |\bq|}  ,\qquad \eta=\del/\ji .
\label{fl}
\end{gather}
Notice that the limit $\eta\to 0$ reduces to the free Hamiltonian in Euclidean space.
The   metric and scalar curvature (\ref{cg}) on the corresponding manifold  $\cM$  turn out to be
\begin{gather}
\dd s^2= \frac{\eta+  |\bq|}{|\bq| }  \dd\bq^2,\qquad
  R=\eta (N-1) \,\frac{\bigl( 4(N-3) r+3(N-2)\eta \bigr)}{ 4r  (\eta+  r)^3},
  \label{fm}
\end{gather}
so that the     domain of $r=|\bq|$ in $\cM$  depends on the sign of $\eta$:
\begin{gather}
\eta>0:\quad r\in  (0,\infty) ;\qquad \eta<0:\quad  r\in [|\eta|,\infty) .
\label{fmm}
\end{gather}
The  intrinsic   potentials (\ref{ch}) are given by
\begin{gather}
{\cal U}_\KC(r)=-\frac{2}{\eta}  \sqrt {\frac{{\eta+  r}}{r}} ,\qquad {\cal U}_\ho(r)= \frac{\eta^2 r} { {4(\eta +  r)}} .
\label{fn}
\end{gather}
Consequently, $ {\cal H}_\eta$ def\/ines two  intrinsic oscillators, which are dif\/ferent systems according to (\ref{fmm}).

It is worth comparing (\ref{fl}) with the Taub-NUT   system~\cite{GM86, FH87, GR88, IK94, IK95, BCJ, JL, GW07}
which can be written as~\cite{annals}:
\begin{gather}
  {\cal H}_{\text{Taub-NUT}}=\frac {\bp^2}{2(1+4m/|\bq|)} +\frac{\mu^2}{2(4m)^2}\left( 1+\frac{4m}{|\bq|}\right)
   \nonumber\\
\phantom{{\cal H}_{\text{Taub-NUT}}}{}  =
  \frac { |\bq| \bp^2 }{2( 4m +   |\bq| )}
+\frac{\mu^2 |\bq|/(4m)^2}{2( 4m + |\bq|  ) } + \frac{\mu^2}{2 |\bq| ( 4m +   |\bq| )}
+\frac{\mu^2 /(4m)}{4m +  |\bq| } .
\label{fo}
\end{gather}
The relationship with $ {\cal H}_\eta $ is established by setting
\[
\eta =4m,\qquad \al=-\frac{\mu^2}{2(4m)^2},
\]
which gives
\begin{gather*}
 {\cal H}_{\eta=4m}+\frac{\mu^2}{(4m)^2} =
  \frac { |\bq| \bp^2 }{2( 4m +   |\bq| )}
+\frac{\mu^2 |\bq|/(4m)^2}{2( 4m + |\bq|  ) }
+\frac{\mu^2 /(4m)}{4m +  |\bq| },
%\label{fp}
\end{gather*}
so that we recover three terms in the ``expanded" expression for  ${\cal H}_{\text{Taub-NUT}}$ (\ref{fo}); namely, the kinetic term def\/ining the geodesic motion on the Taub-NUT space (\ref{fm}), the insintric oscillator potential (\ref{fn})  and the one which comes out  by adding a constant to the oscillator  potential.
There is one missing term, the third one in~(\ref{fo}), which corresponds to  the Dirac  monopole. However we notice that this can be derived from the angular momentum by introducing hyperspherical coordinates in the form~\cite{annals}
\[
\bp^2=p_r^2 + r^{-2} \bL^2\qquad \mbox{and next}\qquad \bL^2\to \bL^2+\mu^2.
\]
From this viewpoint, ${\cal H}_\eta$ can be regarded as an $N$D  MS generalization of the Taub-NUT system which is recovered   for $\eta>0$, being the case with $\eta<0$ a dif\/ferent physical oscillator potential.

 The symmetry properties for ${\cal H}_\eta$ are summarized in

\begin{proposition}\label{proposition7}\qquad
\begin{enumerate}\itemsep=0pt
\item[$(i)$] The   Hamiltonian ${\cal H}_\eta$ \eqref{fl} characterizes two intrinsic oscillator potentials on the corresponding Riemannian space of nonconstant curvature \eqref{fm} according to \eqref{fmm}.

\item[$(ii)$]  ${\cal H}_\eta$ is endowed with the    $(2N-3)$  angular momentum integrals \eqref{bb} together with
 the components of the curved LRL $N$-vector given by   $(i=1,\dots,N)$:
 \begin{gather*}
  \tilde S_{i}=\sum_{k=1}^N p_k \left( q_k p_i - q_i p_k \right)+ \eta\,\frac{ q_i}{|\bq|} {\cal H}_\eta  \qquad \mbox{such that}\quad        \sum_{i=1}^N \tilde S_{i}^2 =    2 \bL^2 ( {\cal H}_\eta-\al) +\eta^2  {\cal H}_\eta^2.
%\label{fj}
\end{gather*}

\item[$(iii)$] The set $\{ {\cal H}_\eta,S^{(m)}, S_{(m)},  \tilde S_{{i}} \}$       $(m=2,\dots,N$ and a fixed index $i)$    is formed by  $2N-1$ functionally independent functions.
\end{enumerate}
\end{proposition}

\section{Outlook and superintegrable  quantization}\label{section5}

\begin{table}[t]
{\footnotesize
 \noindent
\caption{{Maximally superintegrable classical oscillator and KC Hamiltonians in $N$ dimensions.}}
\label{table1}
\medskip
\noindent\hfill
$$
\begin{array}{ll}
\hline
\\[-6pt]
\multicolumn{2}{c}{\mbox {$\bullet$ {Geodesic motion on   Euclidean space}}} \\[4pt]
\multicolumn{2}{c}{  H=\frac 12\, \bp^2 +\al} \\[4pt]
\multicolumn{2}{c} { \mbox { $\ast$ Common $(2N-3)$   angular  momentum integrals of motion} }   \\[4pt]
\multicolumn{2}{c}  {\mbox{ $\displaystyle S^{(m)}= \sum\limits_{1\leq i<j\leq m} (q_ip_j-q_jp_i)^2$,\quad
$\displaystyle S_{(m)}= \sum\limits_{N-m<i<j\leq N} (q_ip_j-q_jp_i)^2$ , \quad $S^{(N)}=S_{(N)}\equiv  \bL^2  $}}  \\[6pt]
{\mbox {$\ast$  ``Seeds'' of the $N$D Fradkin  tensor}}&\quad{\mbox {$\ast$   ``Seeds'' of the LRL $N$-vector}}\\[4pt]
  { S_{ij}=p_ip_j }&\quad{\displaystyle  S_{i}=\sum\limits_{k=1}^N p_k \left( q_k p_i - q_i p_k \right)   }\\[6pt]
\hline
\\[-6pt]
\mbox {$\bullet$ Harmonic oscillator  }&\quad\mbox {$\bullet$ Euclidean KC  }\\[4pt]
 {\displaystyle{ H_U=\frac 12\, \bp^2+\bb \bq^2 +\ga}}&\quad{  \displaystyle{ H_U=\frac 12\, \bp^2+\frac{\del}{| \bq|} +\ji   }}\\[10pt]
 {\mbox {$\ast$   Flat  $N$D Fradkin tensor}}&\quad{\mbox {$\ast$   Flat LRL $N$-vector}}\\
 {  S_{U,{ij}}=p_ip_j  + 2\bb q_i q_j}&\quad{\displaystyle{ S_{U,{i}}=\sum_{k=1}^N p_k \left( q_k p_i - q_i p_k \right) -\del\,\frac{q_i}{|\bq|} }  }\\[12pt]
  \hline
\\[-6pt]
\mbox {$\bullet$ Curved hyperbolic KC   $ (N\ge 3)$}&\quad\mbox {$\bullet$ Curved spherical oscillator   $ (N\ge 3)$ }\\[4pt]
 {\displaystyle{ {\cal H}_\KC=\   \frac{\bp^2}{  2\bq^2}  +\frac{\al}{ \bq^2} }}&\quad{  \displaystyle{ {\cal H}_\ho= \frac 12\, |\bq|  \bp^2  + {\al}{| \bq|}   }}\\[10pt]
  {\mbox {$\ast$   Curved  $N$D Fradkin  tensor}}&\quad{\mbox {$\ast$   Curved LRL $N$-vector}}\\
 {    \tilde S_{ij}=p_ip_j  -2  q_i q_j {\cal H}_\KC}&\quad{\displaystyle{   \tilde S_{i}=\sum_{k=1}^N p_k \left( q_k p_i - q_i p_k \right)+\frac{  q_i}{|\bq|} {\cal H}_\ho  }  }\\[12pt]
 \hline
\\[-6pt]
\mbox {$\bullet$ Darboux III oscillator}&\quad\mbox {$\bullet$ Taub-NUT oscillator} \\[4pt]
{\displaystyle{  {\cal H}_\la = \frac{\bp^2}{2(1+\la \bq^2)}-\frac{\la \al\, \bq^2}{1+\la \bq^2}  }}&\quad{  \displaystyle{  {\cal H}_\eta =  \frac{|\bq| \bp^2}{2(\eta+  |\bq|)}   +\frac{\alpha|\bq|}{\eta+  |\bq|}   }}\\[10pt]
 {\mbox {$\ast$   Curved  $N$D Fradkin  tensor}}&\quad{\mbox {$\ast$   Curved LRL $N$-vector}}\\
 {  \tilde S_{ij}=p_ip_j  -2\la  q_i q_j \left( {\cal H}_\la+\al \right)   }&\quad{\displaystyle{    \tilde S_{i}=\sum_{k=1}^N p_k \left( q_k p_i - q_i p_k \right)+ \eta\,\frac{ q_i}{|\bq|} {\cal H}_\eta }  }\\[12pt]
\hline
\end{array}
$$
\hfill}
\end{table}

  So far we have obtained and interpreted four MS classical Hamiltonian systems on Riemannian spaces of nonconstant curvature by starting from free motion on ${\bf E}^N$ and   applying the St\"ackel transform through the harmonic oscillator and KC potentials.
 The main results here obtained are displayed in Table~\ref{table1} where
   the transition from the ``seeds'' of the Fradkin tensor and the LRL vector up to their curved analogues is laid bare by reading the table through its two columns. Recall, however,   that the Darboux III and the  Taub-NUT oscillators give rise, each of them, to two dif\/ferent physical systems according to the sign of the parameters $\la$ and $\eta$, respectively.

   \begin{table}[t]
{\footnotesize
 \noindent
\caption{{Maximally superintegrable quantum oscillator and KC Hamiltonians in $N$ dimensions.}}
\label{table2}
\medskip
\noindent\hfill
$$
\begin{array}{ll}
\hline
\\[-6pt]
\multicolumn{2}{c} { \mbox { $\ast$ Common $(2N-3)$  quantum  angular  momentum operators} }   \\[4pt]
\multicolumn{2}{c}  {\mbox{  $\displaystyle \hat S^{(m)}= \sum\limits_{1\leq i<j\leq m} (\hat q_i \hat p_j-\hat q_j \hat p_i)^2$,\qquad
$\displaystyle \hat S_{(m)}= \sum\limits_{N-m<i<j\leq N} (\hat q_i \hat p_j-\hat q_j \hat p_i)^2$ , \qquad $\hat S^{(N)}= \hat S_{(N)}\equiv  \hat{\bL}^2  $}}  \\[6pt]
\hline
\\[-6pt]

\mbox {$\bullet$ Quantum hyperbolic KC $ (N\ge 3)$}&\quad\mbox {$\bullet$ Quantum spherical oscillator $ (N\ge 3)$ }\\[4pt]
 {\displaystyle{\hat {\cal H}_\KC=  \frac{1}{  2 \hat\bq^2}\, \hat \bp^2  +\frac{\al}{\hat \bq^2} }}&\quad{  \displaystyle{ \hat{\cal H}_\ho= \frac 12\, | \hat\bq|  \hat\bp^2  + {\al}{| \hat \bq|}   }}\\[10pt]
  {\mbox {$\ast$   Quantum  $N$D Fradkin tensor}}&\quad{\mbox {$\ast$   Quantum LRL $N$-vector}}\\
 {    \hat S_{ij}= \hat p_i \hat p_j  -2 \hat q_i \hat q_j \hat{\cal H}_\KC}&\quad{\displaystyle{     \hat S_{i}=\frac 12 \sum_{k=1}^N \hat p_k \left( \hat q_k \hat p_i -\hat q_i \hat p_k \right)+\frac 12 \sum_{k=1}^N  \left( \hat q_k \hat p_i -\hat q_i \hat p_k \right)\hat p_k +\frac{ \hat q_i}{| \hat\bq|} \hat{\cal H}_\ho  }  } \\

\displaystyle {  \sum_{i=1}^N  \hat S_{ii}= -2 \al }&\quad{\displaystyle{    \sum_{i=1}^N  \hat S_{i}^2=   \hat{\cal H}_\ho^2
-2 \al   \hat \bL^2 -\frac 12 (N-1)^2 \hbar^2 \al } } \\[12pt]

 \hline
\\[-6pt]
\mbox {$\bullet$ Quantum Darboux III oscillator}&\quad\mbox {$\bullet$ Quantum  Taub-NUT oscillator} \\[4pt]
{\displaystyle{\hat  {\cal H}_\la = \frac{1}{2(1+\la \hat \bq^2)}\, \hat \bp^2-\frac{\la \al\, \hat\bq^2}{1+\la \hat\bq^2}  }}&\quad{  \displaystyle{ \hat {\cal H}_\eta =  \frac{| \hat \bq|}{2(\eta+  | \hat\bq|)} \,  \hat \bp^2  +\frac{\alpha| \hat\bq|}{\eta+  | \hat\bq|}   }}\\[10pt]
 {\mbox {$\ast$   Quantum $N$D Fradkin tensor}}&\quad{\mbox {$\ast$   Quantum LRL $N$-vector}}\\
 {  \hat S_{ij}= \hat p_i \hat p_j  -2\la  \hat q_i \hat q_j \left( \hat{\cal H}_\la+\al \right)   }&\quad{\displaystyle{
  \hat S_{i}=\frac 12 \sum_{k=1}^N \hat p_k \left( \hat q_k \hat p_i -\hat q_i \hat p_k \right)+\frac 12 \sum_{k=1}^N  \left( \hat q_k \hat p_i -\hat q_i \hat p_k \right)\hat p_k +\eta\,\frac{ \hat q_i}{| \hat\bq|} \hat{\cal H}_ \eta   }} \\

  \displaystyle {  \sum_{i=1}^N  \hat S_{ii}= 2 \hat  {\cal H}_\la  }&\quad{\displaystyle{    \sum_{i=1}^N  \hat S_{i}^2=    2\hat \bL^2  (\hat {\cal H}_\eta-\al) +\eta^2 \hat {\cal H}_\eta^2  +  \frac 12 (N-1)^2 \hbar^2  (\hat {\cal H}_\eta-\al)     } } \\[12pt]

\hline
\end{array}
$$
\hfill}
\end{table}

   Some related comments are in order. All the Hamiltonians shown in Table~\ref{table1} are constructed on
   spherically symmetric spaces so that they    are endowed with an
     $\frak{so}(N)$  Lie--Poisson symmetry. In particular, let us  consider  the generators of rotations $J_{ij}={q_i}{p_j} - {q_j}{p_i}$ with $i<j$ and $i,j=1,\dots,N$ which span the $\frak
{so}(N)$ Lie--Poisson algebra
\[
\{ J_{ij},J_{ik} \}= J_{jk} ,\qquad  \{ J_{ij},J_{jk} \}= -J_{ik} ,\qquad
\{ J_{ik},J_{jk} \}= J_{ij} , \qquad i<j<k .
\label{ot}
\]
Then   the ``common'' $(2N-3)$ angular momentum integrals $S^{(m)}$ and  $S_{(m)}$  (\ref{bb})  can be written as the
quadratic Casimirs  of  some rotation subalgebras
$\frak {so}(m)\subset \frak {so}(N)$:
\[
S^{(m)}=  \sum_{1\leq i<j\leq m}    J_{ij}^2 ,
\qquad S_{(m)}=  \sum_{N-m<i<j\leq N}   J_{ij}^2 ,
\]
with   $ S^{(N)}=S_{(N)}=\bL^2$ being    the quadratic Casimir of $\frak
{so}(N)$. In this respect, we also notice that all the LRL constants of motion $(S_i,S_{U,i}, \tilde S_i)$ given in Table~\ref{table1} are transformed as $N$-vectors under the action of the generators of $\frak
{so}(N)$ (as it should be):
\[
\{ J_{ij}, \tilde S_k \}= \delta_{ik}\tilde S_j - \delta_{jk}\tilde S_i .
\]

Furthermore, all of  these systems possess an $\frak {sl}(2,\RR)$ coalgebra symmetry as well~\cite{annals}.
If we denote  $J_-=\bq^2$, $J_+= \bp^2$ and    $ J_3=\bq\cdot\bp  $ we have that
\[
\{J_3,J_+\}=2J_+ ,\qquad \{J_3,J_-\}=-2J_- ,\qquad
\{J_-,J_+\}=4J_3,
\]
and the common  integrals  (\ref{bb}) are just the $m$th (left and right) coproducts of the Casimir of $\frak {sl}(2,\RR)$.
This set of $(2N-3)$ integrals is ``universal'' for any Hamiltonian function def\/ined by ${\cal H}={\cal H}(\bq^2,\bp^2,\bq\cdot\bp)$ so that this   always provides, at least, a quasi-MS system~\cite{BH07, sigmaorlando}. Therefore the Hamiltonians shown in Table~\ref{table1} are distinguished systems since they have ``additional'' symmetries.

To end with, we shall present the MS quantization of the four curved classical systems. For this purpose, we remark that the MS quantization of the Darboux III oscillator has been recently obtained in~\cite{pla}, and the corresponding quantum dynamics has been fully solved for $\la>0$ (the case with $\la<0$ is still an open problem). This quantization has been obtained by applying the so called ``Schr\"odinger quantization" procedure~\cite{uwano}, and its relationship with the Laplace--Beltrami and position-dependent-mass quantizations has been established in~\cite{Danilo} by means of similarity transformations (see also~\cite{DaniloSigma}).

Therefore, let us consider
the quantum position and momenta operators, $\hat\bq$, $\hat\bp$, with canonical Lie bracket $
 [\hat q_i,\hat p_j]={\rm i }\hbar \delta_{ij}$. The resulting MS quantum Hamiltonians are summarized in the following f\/inal statement.

\begin{proposition}\label{proposition8}Let ${\cal H}$ be one of the classical Hamiltonians given in Propositions~{\rm \ref{proposition4}--\ref{proposition7}}.
\begin{enumerate}\itemsep=0pt
\item[$(i)$] The Schr\"odinger quantization of ${\cal H}$  and its quantum symmetries are given in Table~{\rm \ref{table2}}.

\item[$(ii)$]  The quantum Hamiltonian $\hat{\cal H}$ is endowed with $(2N-3)$ quantum  angular  momentum ope\-ra\-tors $\hat S^{(m)}$ and  $\hat S_{(m)}$, such that    $\{ \hat{\cal H},\hat S^{(m)}\}$ or
$\{\hat{\cal H},\hat S_{(m)}\}$  $(m=2,\dots,N)$ is  a set of  $N$ algebraically independent  commuting observables.

\item[$(iii)$]  If $\hat{\cal H}\in(\hat {\cal H}_\KC, \hat {\cal H}_\la)$ then it commutes with the $N^2$ components, $\hat S_{ij}$, of a quantum Fradkin tensor $(i,j=1,\dots,N)$. The set $\{\hat{\cal H} ,\hat S^{(m)}, \hat S_{(m)},  \hat S_{{ii}} \}$       $(m=2,\dots,N$ and a fixed index $i)$    is formed by  $2N-1$ algebraically independent  commuting observables.

\item[$(iv)$] When  $\hat{\cal H}\in(\hat {\cal H}_\ho, \hat {\cal H}_\eta)$ this commutes with the $N$ components, $\hat S_{i}$, of a quantum LRL vector $(i=1,\dots,N)$. The set $\{\hat{\cal H} ,\hat S^{(m)}, \hat S_{(m)},  \hat S_{{i}} \}$       $(m=2,\dots,N$ and a fixed index $i)$    is constituted  by  $2N-1$ algebraically independent  commuting observables.
\end{enumerate}
\end{proposition}

\subsection*{Acknowledgements}

 This work was partially supported by the Spanish MICINN   under grants    MTM2010-18556   and FIS2008-00209, by the    Junta de Castilla y Le\'on  (project GR224), by the Banco Santander--UCM  (grant GR58/08-910556)
   and by  the Italian--Spanish INFN--MICINN (project ACI2009-1083). F.J.H.\ is deeply grateful to W.~Miller~Jr.\ for very  helpful suggestions on the St\"ackel transform as well on superintegrability.

\pdfbookmark[1]{References}{ref}
\LastPageEnding

\end{document}